\documentclass[twocolumn]{aastex63}
\usepackage{inputenc}
\usepackage{comment}
\usepackage{colortbl}
\usepackage{hyperref}
\usepackage{graphicx}

\received{28 April 2023}
\revised{10 July 2023}
\accepted{14 July 2023}
\submitjournal{ApJ}

\shorttitle{4FGL Pulsar Follow-Up}
\shortauthors{Kerby et al. 2022}

\begin{document}

\title{\textit{Swift} Follow-Up of Reported Radio Pulsars at \textit{Fermi} 4FGL Unassociated Sources}

\author[0000-0003-2633-2196]{Stephen Kerby}
\affiliation{Department of Astronomy and Astrophysics \\
 Pennsylvania State University,
University Park, PA 16802, USA}

\author[0000-0002-5068-7344]{Abraham D. Falcone}
\affiliation{Department of Astronomy and Astrophysics \\
 Pennsylvania State University,
University Park, PA 16802, USA}

\author[0000-0002-5297-5278]{Paul S. Ray}
\affiliation{Space Science Division, U.S. Naval Research Laboratory, Washington, DC 20375, USA}

\begin{abstract}

Following the discovery of radio pulsars at the position of \textit{Fermi}-LAT unassociated sources by the TRAPUM group, we conduct \textit{Swift}-XRT observations of six of those 4FGL sources to determine if any pulsar-like X-ray sources are present and to confirm the reported detection of an X-ray counterpart via eROSITA at 4FGL J1803.1$-$6708. At two of the six targets, we detect no X-ray sources at the TRAPUM radio position, placing an upper limit on the $0.3-10.0$ keV flux. At 4FGL J1803.1$-$6708 we find an X-ray source at the TRAPUM and eROSITA position. At 4FGL J1858.3$-$5424 we find a new X-ray counterpart at the TRAPUM position with S/N=4.17, but also detect a distinct and separate X-ray source. At 4FGL J1823.8$-$3544 and 4FGL J1906.4$-$1757 we detect no X-ray flux at the TRAPUM positions, but we do detect separate X-ray sources elsewhere in the \textit{Fermi} error ellipse.  At these last two targets, our newly detected \textit{Swift} sources are possible alternatives to the radio pulsar associations proposed by TRAPUM. Our findings confirm several of the discoveries reported by the TRAPUM group but suggest that further observations and investigations are necessary to confirm the low-energy counterpart of several unassociated sources.

\end{abstract}

\keywords{neutron stars --- pulsars}

\section{Introduction}
\label{sec:Intro}

The \textit{Fermi}-LAT 4FGL-DR3 catalog contains over six thousand sources, mostly extragalactic blazars or nearby gamma-ray pulsars \citep{Abdollahi2020,Ballet2020,FermiDR32022}.  Across the three iterations of the catalog, approximately 2000 4FGL sources are considered ``unassociated", lacking confident astronomical classification or counterparts. Given that a significant fraction of the associated sources are pulsars in the \textit{Fermi}-LAT gamma-ray pulsar catalog \citep{Abdo2013}, it is feasible that many of the unassociated sources are also gamma-ray pulsars.

The Neil Gehrels \textit{Swift} Observatory (aka \textit{Swift}) \citep{Gehrels2004}, originally designed for pursuing gamma-ray bursts, has in recent years conducted an ongoing campaign of observations near the \textit{Fermi} unassociated sources, aiming to find X-ray counterparts and localize the astronomical sources of gamma-ray emission with much greater angular precision than the 4FGL uncertainty ellipses. With the XRT \citep{Burrows2005} and UVOT \citep{Roming2005} instruments onboard \textit{Swift} conducting X-ray observations from $0.3 - 10.0 \:\rm{keV}$ and UV-visual observations from $450 - 900 \:\rm{nm}$, \textit{Swift}'s observations have discovered hundreds of possible counterparts to the unassociated sources \citep{Kaur2019,Kerby2021,Kerby2021b}. Using machine learning classification to compare the combined Gamma-ray/X-ray/UV/optical spectral features of the unassociated sources to known gamma-ray pulsars and blazars, hundreds of the unassociated sources have been classified as likely pulsars or blazars.

In \cite{TRAPUM2022}, the TRAPUM consortium using the MeerKAT telescope \citep{Jonas2009} announced the discovery of several radio pulsars after a L-band radio survey of \textit{Fermi} LAT unassociated sources, including new redback/black widow pulsars. \cite{TRAPUM2022} also notes one target, 4FGL J1803.1$-$6708, where eROSITA X-ray data indicates significant X-ray flux at the radio position. To follow up at the TRAPUM radio positions, we submitted Target of Opportunity (ToO) requests to \textit{Swift} for six of the 4FGL unassociated sources in question to search for X-ray counterparts. \textit{Swift}-XRT and -UVOT observations at the 4FGL targets in question can provide localization and multi-wavelength characterization of any X-ray sources in the regions of the gamma-ray emission.

In this work we investigate the 4FGL unassociated regions noted as hosting pulsars by the TRAPUM consortium.  In \S\ref{sec:Obs} we describe our target selection and the \textit{a priori} expectations for X-ray followup at suspected pulsars. In \S\ref{sec:DataAna} we discuss our analysis of \textit{Swift}-XRT and -UVOT observations, plus we apply previously used classification routines to the pulsars as independent checks on previous works. In \S\ref{sec:Results} we discuss our results at each target and compare with the TRAPUM results. In \S\ref{sec:Coin} we provide a summary of our findings and propose next steps for the continued investigation of pulsars at 4FGL unassociated sources.

\section{Targets and Observational Expectations} 
\label{sec:Obs}

Of the seven 4FGL sources associated with announced TRAPUM pulsar discoveries, we selected six for follow-up observations with \textit{Swift}. Prior to our observational campaign, three of the six selected targets had less than $3 \:\rm{ks}$ of previous \textit{Swift}-XRT exposure, allowing for our additional observations to dramatically expand coverage at those sites. 4FGL J1858.3$–$5424 and 4FGL J1803.1$-$6708 had low-S/N X-ray counterparts already present after $\sim 3 \:\rm{ks}$ of observations; our additional observations could perhaps increase the S/N of these tentative counterparts into firm detections. Finally, 4FGL J1623.9$–$6936 was selected as a target despite already having $\sim 5 \:\rm{ks}$ of \textit{Swift} observations due to its proximity with the other sources. All six targets provided an opportunity to independently investigate the high-energy emission from detected radio pulsars. 

At each target, we initially requested $10000$ seconds of observations in photon counting (PC) mode.  Because the neural network classifier (NNC) built in \cite{Kerby2021b} uses UVOT V-band magnitudes for pulsar-blazar classification, we furthermore requested that simultaneous \textit{Swift}-UVOT exposures be in the V-band.  Table \ref{tab:obs} lists the targets and exposure times that facilitated this analysis. New observations were combined with all archival \textit{Swift} exposures available on HEASARC for each 4FGL source. 

\begin{table*}
\caption{The 4FGL target sources of this work, including the \textit{Swift} target ID for reference, the total \textit{Swift}-XRT exposure, and the number of expected unrelated but coincident X-ray interlopers above $S/N=3$ and $=4$ thresholds}
\label{tab:obs}
\centering
 \begin{tabular}{lcccccc}
 \hline
 4FGL Source & RA & Dec & Target ID & XRT Exposure &  Num. Expected & Num. Expected \\
 & (J2000) & (J2000) & & (ks) & S/N $> 3$ & S/N $> 4$ \\
 \hline\hline
J1623.9$-$6936 & 245.9865 & -69.6069 & 14886 & 17.5 & 0.34 & 0.17 \\
J1757.7$-$6032 & 269.4489 & -60.5374 & 47265 & 12.6 & 0.06 & 0.02 \\
J1803.1$-$6708 & 270.7910 & -67.1347 & 84818 & 13.7 & 0.15 & 0.07 \\
J1823.8$-$3544 & 275.9677 & -35.7402 & 14436 & 14.3 & 0.50 & 0.25 \\
J1858.3$-$5424 & 284.5760 & -54.4123 & 14438 & 35.2 & 1.05 & 0.55 \\
J1906.4$-$1757 & 286.6107 & -17.9509 & 14437 & 12.4 & 0.35 & 0.17 \\
 \hline
 \end{tabular}
\end{table*}

When searching for X-ray counterparts in a gamma-ray source uncertainty ellipse, it is worthwhile to evaluate the expected number of coincidental X-ray sources expected in the uncertainty ellipse, and to evaluate whether the detection of X-ray emission from a pulsar is likely at all. By finding the expected number of X-ray sources in each region from a density function, we can estimate the likelihood that a detected X-ray source is unrelated to, but spatially coincident with, a \textit{Fermi} uncertainty ellipse.  If the expected number of $S/N > 4$ sources is less than one, we can have reasonable confidence that a detected X-ray source in the 4FGL uncertainty ellipse is related to the gamma-ray emission.

To estimate the expected number of X-ray sources, we use a selection of \textit{Swift}-XRT fields near gamma-ray burst positions. As these fields are distributed randomly across the sky and have a single source that can be strongly attributed to the GRB, any additional X-ray sources in the fields can be used to calculate spatial density of randomly detected X-ray sources.  Table \ref{tab:obs} lists the expected number of unrelated but coincident X-ray sources in the 4FGL uncertainty ellipses, derived from \textit{Swift}'s extensive GRB observations. We show the estimated number of coincident sources for $S/N=3$ (a marginal \textit{Swift} detection) and $S/N=4$ (a confident \textit{Swift} detection), two levels characteristic of counterparts to other unassociated source.

\subsection{Observational Expectations}
\label{sec:Median}

Given that the TRAPUM consortium has detected radio pulsations near several 4FGL targets, what is the likelihood of detecting an X-ray excess from such a pulsar with \textit{Swift}-XRT observations? To examine this question, we used the sample of 71 known gamma-ray pulsars with X-ray fluxes from \cite{Kerby2021b} to predict the $S/N$ ratio for a gamma-ray pulsar that can be detected both by \textit{Swift}-XRT and \textit{Fermi}-LAT. This sample is based off the second \textit{Fermi} pulsar catalog \citep{Abdo2013}, using only gamma-ray pulsars with detected X-ray fluxes, so it is biased towards gamma-ray and X-ray bright pulsars. Given that the pulsars in the unassociated sources are generally dimmer in gamma-rays than the known pulsars in \cite{Abdo2013}, our estimation of pulsar $S/N$ is an optimistic upper limit on the chances of detecting pulsars with further X-ray observations. Furthermore, because the pulsar sample built in \cite{Kerby2021b} contains only those pulsars with X-ray fluxes, the pulsar/blazar classification pursued below for the TRAPUM pulsars is similarly limited.

\begin{figure}
    \centering
    \includegraphics[width=\columnwidth]{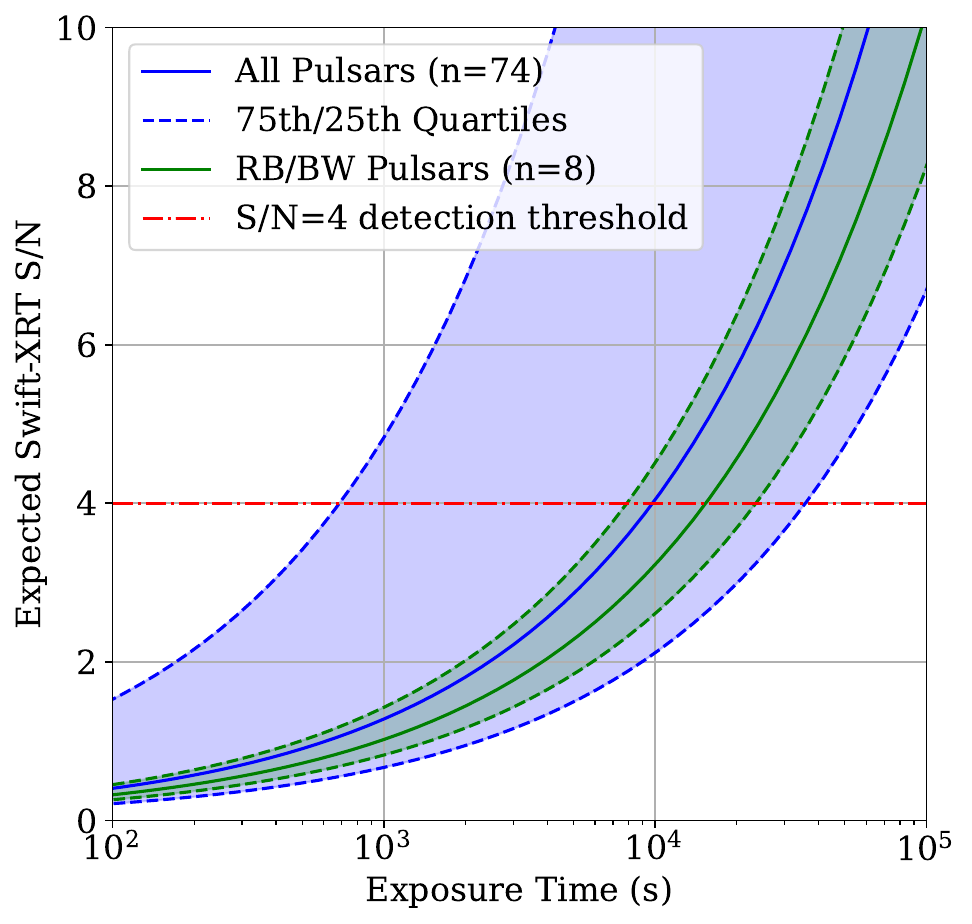}
    \caption{Median and quartile $S/N$ curves for known pulsars with X-ray and V-band fluxes (blue) and a subsample of known redback/black widow millisecond pulsars (green). With a $10 \:\rm{ks}$ exposure with \textit{Swift}-XRT, half of pulsars and a quarter of redback/black widows exceed a $S/N = 4$ detection threshold.}
    \label{fig:SN}
\end{figure}

Taking the median X-ray flux and power law X-ray photon index of the sample of known pulsars ($F_X = 8.49 \times 10^{14} \:\rm{erg/s/cm^2}$ and $\Gamma_X = 1.77$), we use \verb|WebPIMMS| to predict Swift-XRT PC-mode count rate with galactic hydrogen column density $n_H = 0$ (the resulting count rate is only slightly modified for high $n_H$). We find that the median pulsar in the sample has a count rate of $r_s = 1.78 \times 10^{-3} \:\rm{/s}$ with \textit{Swift}-XRT. To compute expected background, we use the $0.3-10.0 \:\rm{keV}$ background count rate of $r_b = 0.45 \times 10^{-3} \:\rm{/s/\square\arcmin}$ from \cite{2013Delia} in a source region of radius $20 \arcsec$, twice the half-power radius of a \textit{Swift} point source. Assuming Poisson arrival statistics, the $S/N$ for an observation in a region of size $A=\pi R^2$ after an exposure time $t$ is given by

$$ S/N = \frac{r_s t}{\sqrt{(r_s+r_bA)t}} $$

\noindent with which we compute the expected $S/N$ for \textit{Swift}-XRT observations up to $100 \:\rm{ks}$. We repeat this process for the smaller subsample of the eight known redback and black widow pulsars in our pulsar list using the median X-ray flux and photon index of that subsample ($F_X = 4.24 \times 10^{14} \:\rm{erg/s/cm^2}$ and $\Gamma_X = 2.35$). This subsample of redback and black widow pulsars is more similar to the millisecond pulsars detected in \cite{TRAPUM2022}, but its small size and the biases of our larger pulsar sample once again preclude strong conclusions from this comparison alone. We plot the $S/N$ for these median pulsars and redback/black widows, plus the 25th and 75th percentiles of each sample, in Figure \ref{fig:SN}.

The expected $S/N$ curves show that even after $10 \:\rm{ks}$ of \textit{Swift}-XRT observations, at best only half of gamma-ray pulsars would be detected as a $S/N \ge 4$ X-ray source. Though the redback/black widow subsample used is small, it suggests that only a quarter of redback/black widow pulsars would reach that detection threshold after $10 \:\rm{ks}$, though half of such pulsars would reach $S/N = 3$.  Therefore, Figure \ref{fig:SN} also highlights that the millisecond pulsars noted by the TRAPUM consortium might not be bright enough in X-rays to be detected by our \textit{Swift}-XRT observations. If that is the case, it would not be surprising if we detect no X-ray sources in the uncertainty ellipses of some of the 4FGL targets hosting TRAPUM pulsars.

This calculation is an optimistic estimate of gamma-ray pulsar detection for two primary reasons. First, the gamma-ray pulsars of the \textit{Fermi} pulsar catalog are generally brighter in gamma-rays than the unassociated sources. If the X-ray to gamma-ray flux ratio of pulsars is consistent between these two samples, the expected X-ray flux from a pulsar in an unassociated source will be likewise smaller than the fluxes used to calculate the curves in Figure \ref{fig:SN}. Secondly, the S/N estimates above were created specifically using a sample of pulsars detected in X-rays and gamma-rays, meaning the gamma-ray pulsars that are systematically dimmer in X-rays are excluded from the sample above.

There may be an X-ray source in the 4FGL uncertainty ellipse that is spatially separate from the radio positions of the TRAPUM discoveries. In this case, the X-rays may be originating from another astronomical object, perhaps a background blazar or another pulsar, and the true source of the 4FGL gamma-ray emission would be uncertain without further study.

\section{Data Analysis} \label{sec:DataAna}

\subsection{\textit{Swift}-XRT Observations}

Upon downloading the new and archival \textit{Swift} observations, we reprocessed and cleaned all PC-mode level 1 event files using \verb|xrtpipeline| v.0.13.5 from the HEASOFT software\footnote{\url{https://heasarc.gsfc.nasa.gov/docs/software.html}}.  We merged the events and exposure files for each 4FGL target using \verb|xselect| v.2.4g and \verb|ximage| v.4.5.1, resulting in a single summed event list, exposure map, and ancillary response file for each source. In this research we only used Photon Counting (PC) mode observations. PC-mode observations preserve 2D locations of arriving photons and are suited for dim sources like pulsars and counterparts to unassociated sources.

We used the \verb|ximage| routines \verb|detect| and \verb|SOSTA| to search for X-ray sources within each uncertainty ellipse, as well as specifically at the position of each radio source discovered by the TRAPUM group. We conducted X-ray analysis on all detected X-ray sources, not just those at TRAPUM positions. The source region for each position is centered on the X-ray position with a radius of 8 pixels (20 arcseconds). The background regions were annular in shape with inner radius 20 pixels (47 arcseconds) and outer radius 60 pixels (141 arcseconds). TRAPUM positions with no \textit{Swift}-XRT source were instead analyzed for upper limits on X-ray flux.

For detected X-ray sources, we used \verb|Xspec| v.12.10.1f \citep{Arnaud1996} to fit each spectrum with a simple power-law fit. The model nested three functions: \verb|tbabs|, \verb|cflux|, and \verb|powerlaw|. \verb|cflux| calculated the total unabsorbed flux between 0.3 and 10 keV while \verb|tbabs| modeled line-of-sight hydrogen absorption using the fixed galactic values from the \verb|nH| lookup function \citep{Kalberla2005}. Fitting was executed using the C-statistic as the optimization metric.

\begin{figure*} \label{fig:fields}
\centering
\begin{tabular}{cc}
 \includegraphics[height=5.5cm]{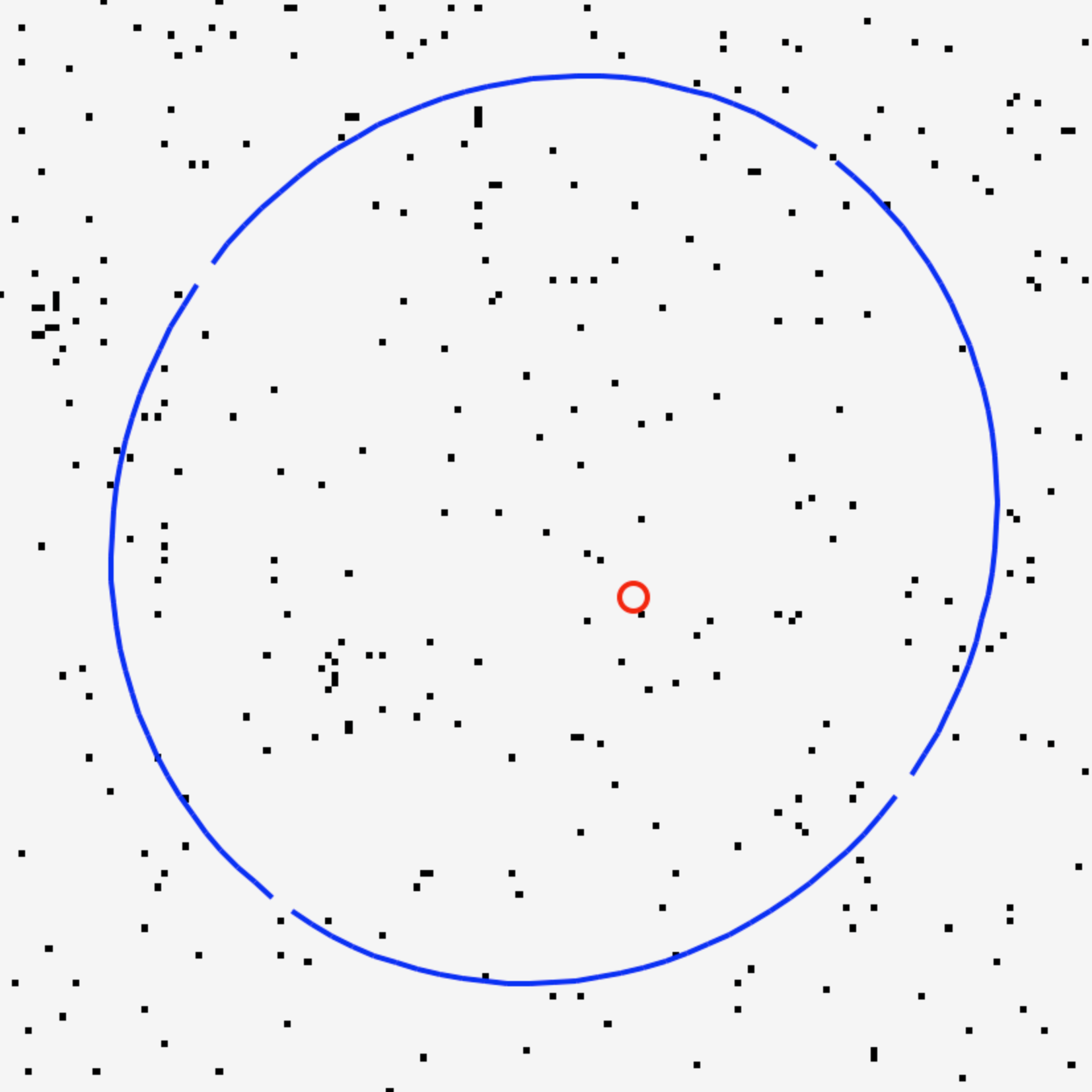} & \includegraphics[height=5.5cm]{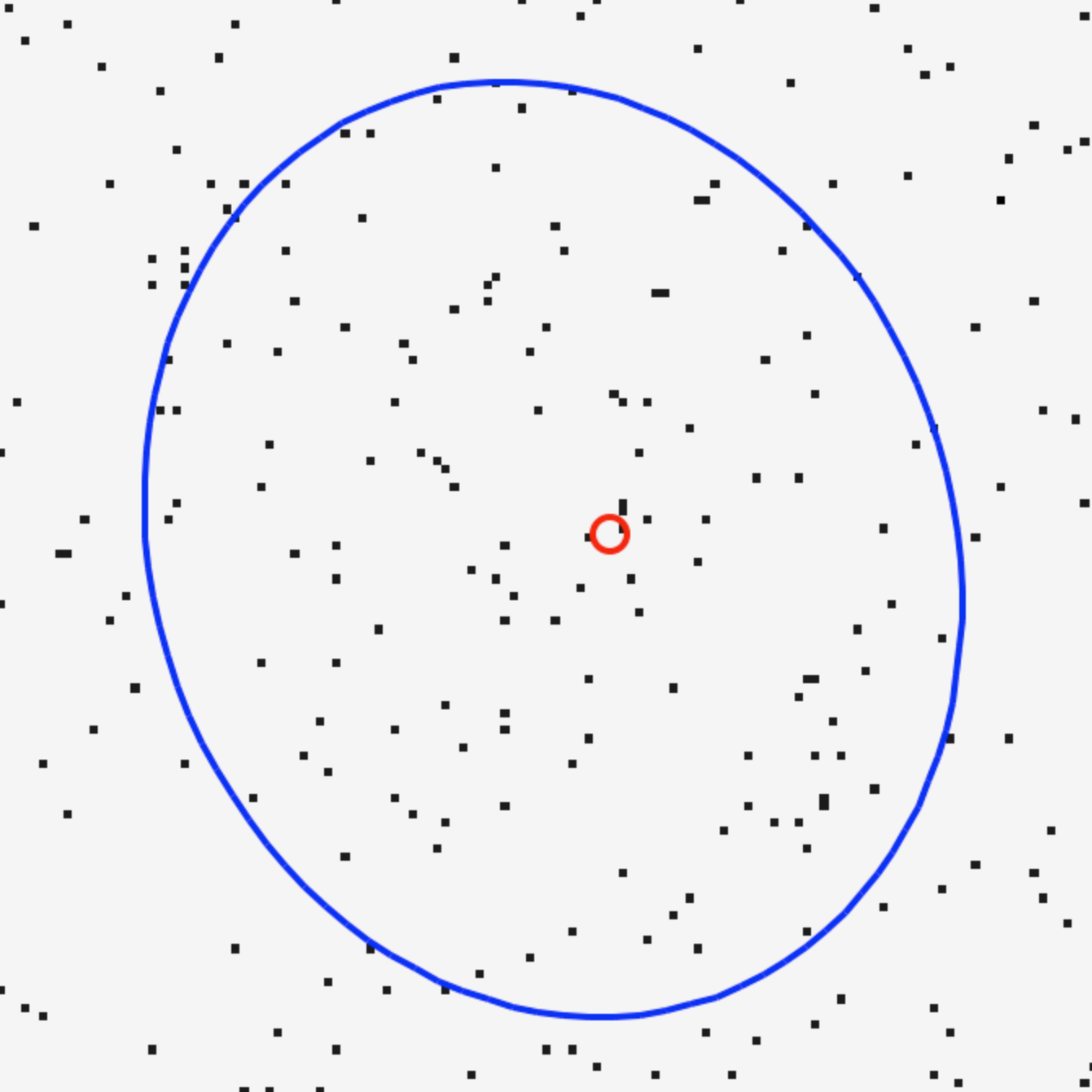} \\
(a) 4FGL J1623.9$-$6936 & (b) 4FGL J1757.7$-$6032  \\[6pt]
 \includegraphics[height=5.5cm]{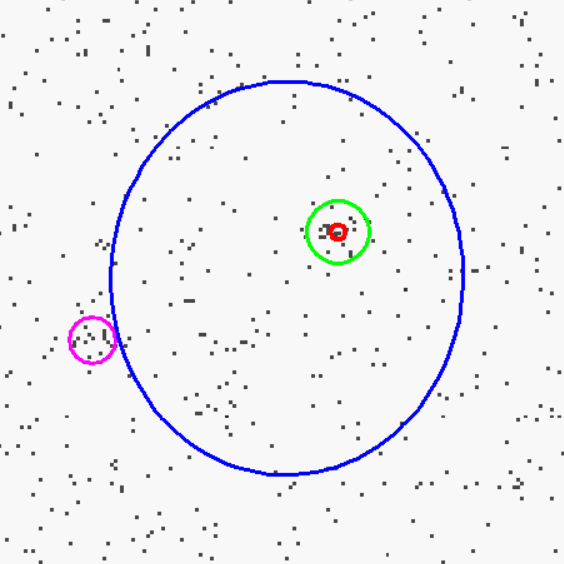} & \includegraphics[height=5.5cm]{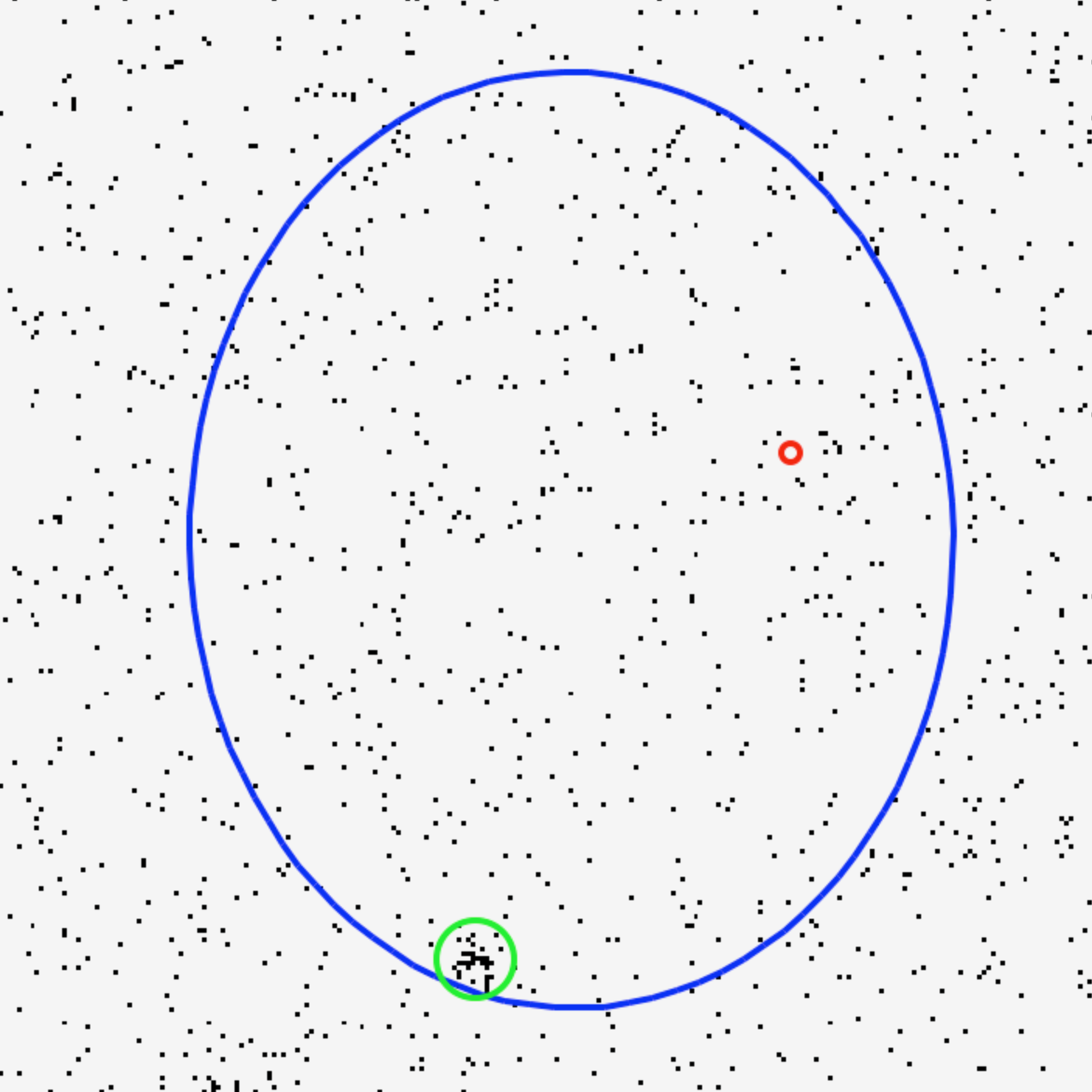} \\
(c) 4FGL J1803.1$-$6708 & (d) 4FGL J1823.8$-$3544 \\[6pt]
 \includegraphics[height=5.5cm]{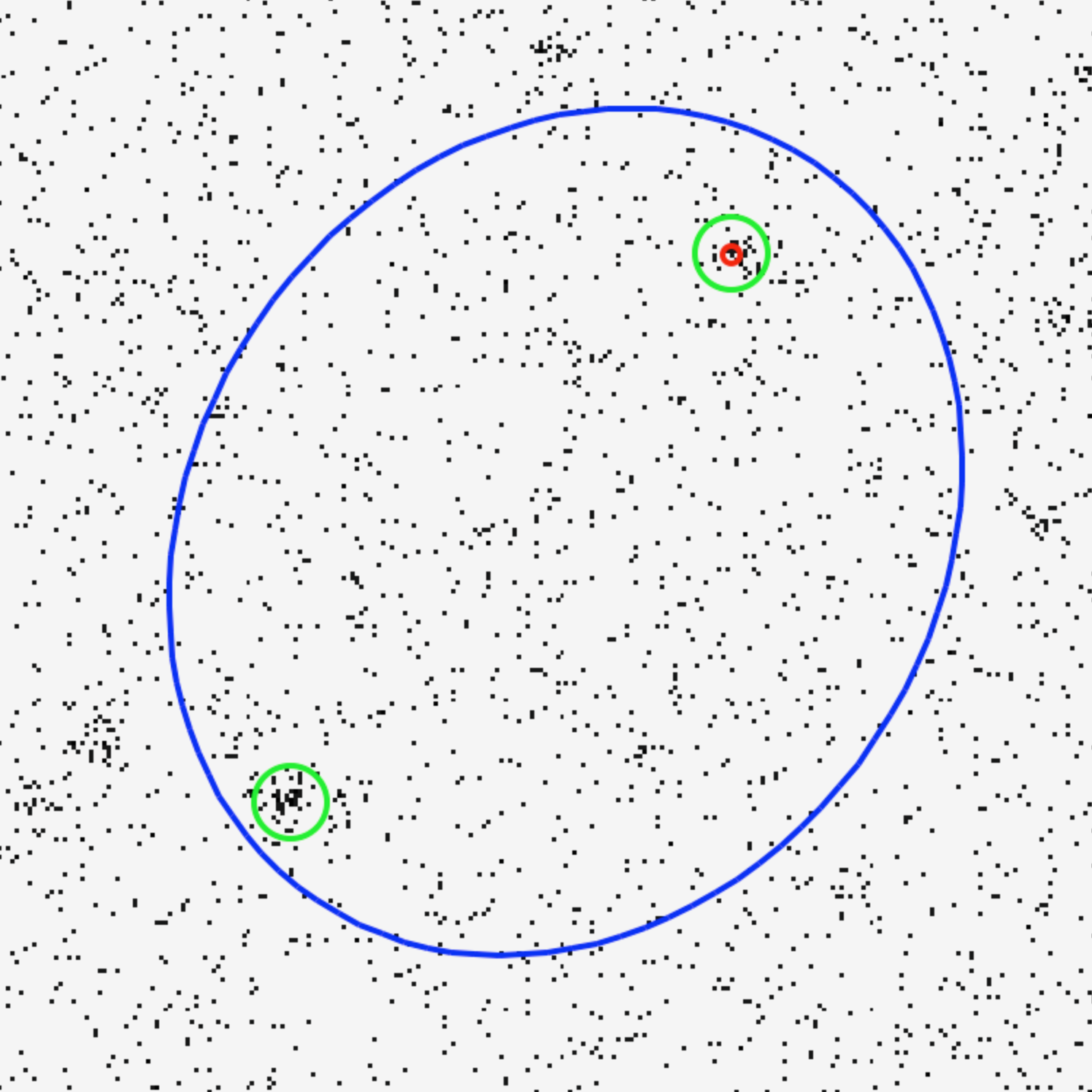} & \includegraphics[height=5.5cm]{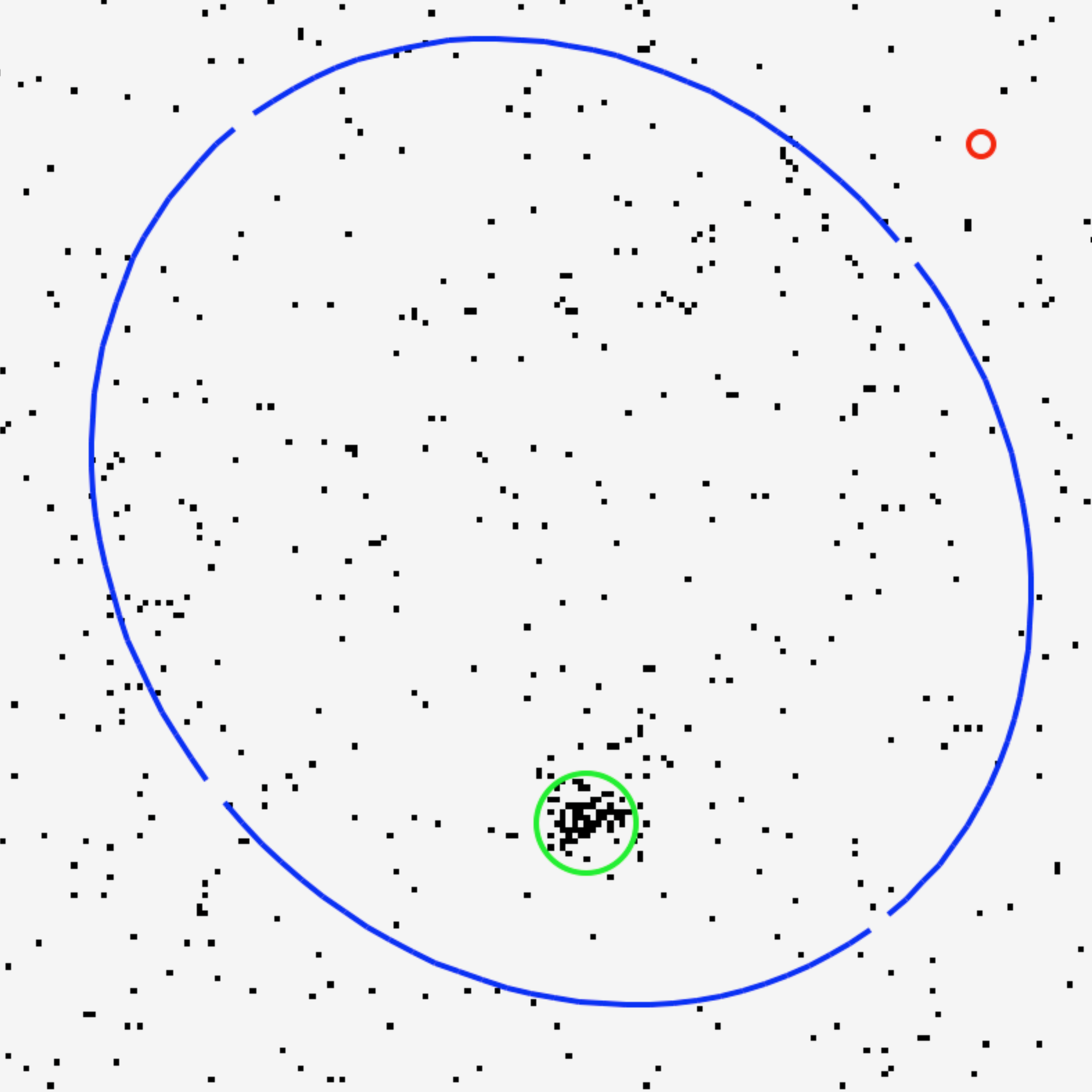} \\
(e) 4FGL J1858.3$-$5424 & (f) 4FGL J1906.4$-$1757 \\[6pt]
\end{tabular}
\caption{\textit{Swift}-XRT fields the six 4FGL targets (95\% uncertainty ellipses in blue) with TRPAUM source position (red) and XRT source regions (green, $r=20\arcsec$). For J1803.1$-$6708, we also include in magenta the secondary X-ray source outside the 4FGL ellipse.}
\end{figure*}

\subsection{\textit{Swift}-UVOT V-band Observations}

For each detected X-ray source, we analyzed the simultaneous \textit{Swift}-UVOT exposures to characterize the optical brightness of possible counterparts in the \textit{Swift} V band \footnote{Henceforth, V band; not to be confused with the similar Johnson V band centered at $551.0 \:\rm{nm}$}.  In \cite{Kerby2021b}, we found that V-band observations are a powerful discriminator between blazars which are bright at all energy ranges and pulsars which are dim in optical light.  To obtain V-band magnitudes, we summed all available V-band UVOT observations for each 4FGL target using \verb|uvotimsum|; as some archival \textit{Swift} observations were not in the V-band, the summed exposure time in UVOT V-band is different in length from the summed \textit{Swift}-XRT exposure.

Next, we search for UVOT sources within $5 \arcsec$ of each XRT source.  Several XRT sources lacked any UVOT source in their summed exposure maps; for these sources, we left the UVOT source region at the XRT position to obtain an upper limit on V-band flux.  Finally, we selected a nearby empty circular region of radius at least $10 \arcsec$ to serve as a background for UVOT analysis.  Using these regions, we obtained V-band AB magnitudes using \verb|uvotsource|. These magnitudes are listed in Table \ref{tab:results}, at the end of this paper. 

Figure 2 shows the \textit{Swift}-XRT fields for each target, including the \textit{Fermi} unassociated 95\% uncertainty ellipse (blue), the \textit{Swift}-XRT source location (green), and the TRAPUM radio position (red).
 
\subsection{Neural Network Classification}
\label{sec:NNC}

After obtaining spectral parameters from the \textit{Swift}-XRT and -UVOT observations, we adapted the NNC developed in \cite{Kerby2021b} to determine whether the newly-discovered X-ray/UVOT sources are pulsar-like in nature. If the new sources are pulsar-like, they could be candidate links between the gamma-ray emission of the \textit{Fermi}-LAT 4FGL catalog and the radio pulsars detected by the TRAPUM consortium.  The NNC described in \cite{Kerby2021b} used a training sample of 71 known pulsars and 635 known blazars with catalogued gamma-ray, X-ray, and optical spectral features to classify unassociated sources into likely blazars or pulsars. 

While the 4FGL catalog contains some objects that are neither pulsars nor blazars, the numerical majority of associated sources are either galactic pulsars or extragalactic blazars, so the classifier in \cite{Kerby2021b} was designed with binary probabilistic classification in mind. It achieved high levels of accuracy and confidence when tested on a variety of validation subsamples, while returning inconclusive values for blazar probability $P_{bzr}$ for sources that were later independently determined to be neither pulsar nor blazar.

In this work, we apply the classifier developed in \cite{Kerby2021b} to the newly detected counterparts to directly test the classifier on independently discovered pulsars. A result of low $P_{bzr}$ scores for the sources examined here would independently confirm the capability of the classifier to detect pulsars, and high $P_{bzr}$ scores would suggest that the neural network must be treated skeptically in its classifications. Inconclusive results would suggest that pulsars at the unassociated sources may behave different from the brighter pulsars in the training dataset. Using the classifier also enables comparisons between the pulsars examined here and brighter gamma-ray and X-ray pulsars.

To keep as many parameters as possible distance-independent, we use X-ray and V-band fluxes in ratio with gamma-ray flux; because the unassociated sources have lower \textit{Fermi}-LAT flux than the known gamma-ray pulsars and blazars used to train the NNC, having distance-independent properties maintains close correspondence between the unassociated sources and our training dataset. Because we take the logarithm of the ratio between V-band flux to gamma-ray flux, and because our ML procedure first rescales and recenters each parameter, the exact reference flux and magnitude used to convert $m_V$ to $F_V$ are irrelevant. For our purposes, we adopt the conversion from \cite{Bessell1998}.

$$ \log{F_V} = -\frac{m_V + 21.1}{2.5} $$

The parameters for each source in the training sample of the NNC are:
\begin{itemize}
    \item X-ray photon index, $\Gamma_X$
    \item Gamma-ray photon index, $\Gamma_\gamma$ (\verb|PL_Index| in the 4FGL catalog)
    \item The logarithm of gamma-ray flux, $\log{F_\gamma}$, in $\rm{erg/s/cm^2}$ (\verb|Energy_Flux| in the 4FGL catalog)
    \item The logarithm of X-ray to gamma-ray flux ratio, $\log{F_X/F_\gamma}$
    \item The logarithm of V-band to gamma-ray flux ratio, $\log{F_V/F_\gamma}$.
    \item The significance of the curvature in the gamma-ray spectrum (\verb|PLEC_SigCurv| in the 4FGL catalog)
    \item The year-over-year gamma-ray variability index (\verb|Variability_Index| in the 4FGL catalog).
\end{itemize}

Converting the spectral parameters of the new X-ray sources to these formats, we insert each into the NNC and obtain $P_{bzr}$ values that describe the probability that each detection is a blazar; in this binary classification, $P_{psr} = 1 - P_{bzr}$. The $P_{bzr}$ results of running the NNC on the new X-ray sources are shown in Table \ref{tab:results}.

\section{Results}
\label{sec:Results}

\subsection{Individual Target Discussion}

\subsubsection{J1623.9-6936}

In the error ellipse of 4FGL J1623.9$-$6936 we detect no X-ray source at the TRAPUM radio position or within the 4FGL 95\% uncertainty ellipse. A dim source is present just outside the ellipse, but we do not view that detection as a likely alternate counterpart. Using \verb|SOSTA|, we establish a $3 \sigma$ upper limit on the $0.3-10.0 \:\rm{keV}$ count rate of $1.08 \times 10^{-3}$ counts per second at the radio position. Assuming a typical absorbed power law spectrum with slope $\Gamma_X = 2$, this count rate limit corresponds to a \textit{Swift}-XRT X-ray flux upper limit of $3.7 \times 10^{-14} \:\rm{erg/s/cm^2}$ estimated by \verb|webPIMMS|.

\subsubsection{J1757.7-6032}

In the error ellipse of 4FGL J1757.7$-$6032 we find a marginal $\rm{S/N} \approx 1.6$ excess at the TRAPUM radio position. \verb|SOSTA| provides a $3 \sigma$ upper limit on the $0.3-10.0 \:\rm{keV}$ count rate of $2.5 \times 10^{-3}$ counts per second. Via \verb|webPIMMS| and assuming $\Gamma_X = 2$ as above, this count rate limit corresponds to a \textit{Swift}-XRT upper limit X-ray flux of $8.6 \times 10^{-14} \:\rm{erg/s/cm^2}$.

\subsubsection{J1803.1-6708}

In the error ellipse of 4FGL J1803.1$-$6708, we detect a $\rm{S/N} = 4.25$ X-ray source at the same position as the TRAPUM radio pulsar. This X-ray source is spatially coincident with the eROSITA X-ray source noted in \cite{TRAPUM2022}, confirming that detection. Using the procedure described in section \ref{sec:DataAna}, we conduct a power-law fit to the X-ray data at this location. Using the NNC from \cite{Kerby2021b}, we find that this source has $P_{bzr} = 0.145$, suggesting a moderately pulsar-like source.  Furthermore, \cite{TRAPUM2022} obtained a timing solution with gamma-ray pulsations from 4FGL J1803.1$-$6708, giving timing evidence that this source is indeed a radio and gamma-ray pulsar

Additionally, \verb|DETECT| notes a dimmer X-ray source outside the western edge of the \textit{Fermi} uncertainty ellipse, in magenta in Figure 2c. This source has $\rm{S/N} = 3.0$, a less significant detection compared to the primary counterpart at the TRAPUM position.  Given the TRAPUM radio position, the reported gamma-ray timing solution for this source \citep{TRAPUM2022}, and coincidental eROSITA and \textit{Swift} X-ray detections at the primary X-ray source, it is very unlikely that this secondary X-ray detection is the counterpart to the unassociated source.

\subsubsection{J1823.8-3544}

At 4FGL J1823.8$-$3544 we detect no X-ray source at the TRAPUM radio position.  \verb|SOSTA| gives a $3 \sigma$ upper limit on the $0.3-10.0 \:\rm{keV}$ count rate of $1.25 \times 10^{-3}$ counts per second at the TRAPUM position. Via \verb|webPIMMS| and assuming $\Gamma_X = 2$ as above, this count rate limit corresponds to a upper limit \textit{Swift}-XRT X-ray flux of $4.3 \times 10^{-14} \:\rm{erg/s/cm^2}$.

\textit{Swift}-XRT does detect a fairly bright $\rm{S/N} \approx 4.95$ X-ray source in the southern part of the 4FGL uncertainty ellipse.  This X-ray source has two UVOT sources within $5 \arcsec$ of its position, making it difficult to estimate the exact UV/optical counterpart to the X-ray source. One of the UVOT counterparts has $m_V = 16.98$ while the other has $m_V = 17.19$, very close in magnitude, so for the purposes of inputting a magnitude into the machine learning classifier from \cite{Kerby2021b} we adopt $m_V \approx 17.1$ for this source.  

\begin{figure}
    \centering
    \includegraphics[width=\columnwidth]{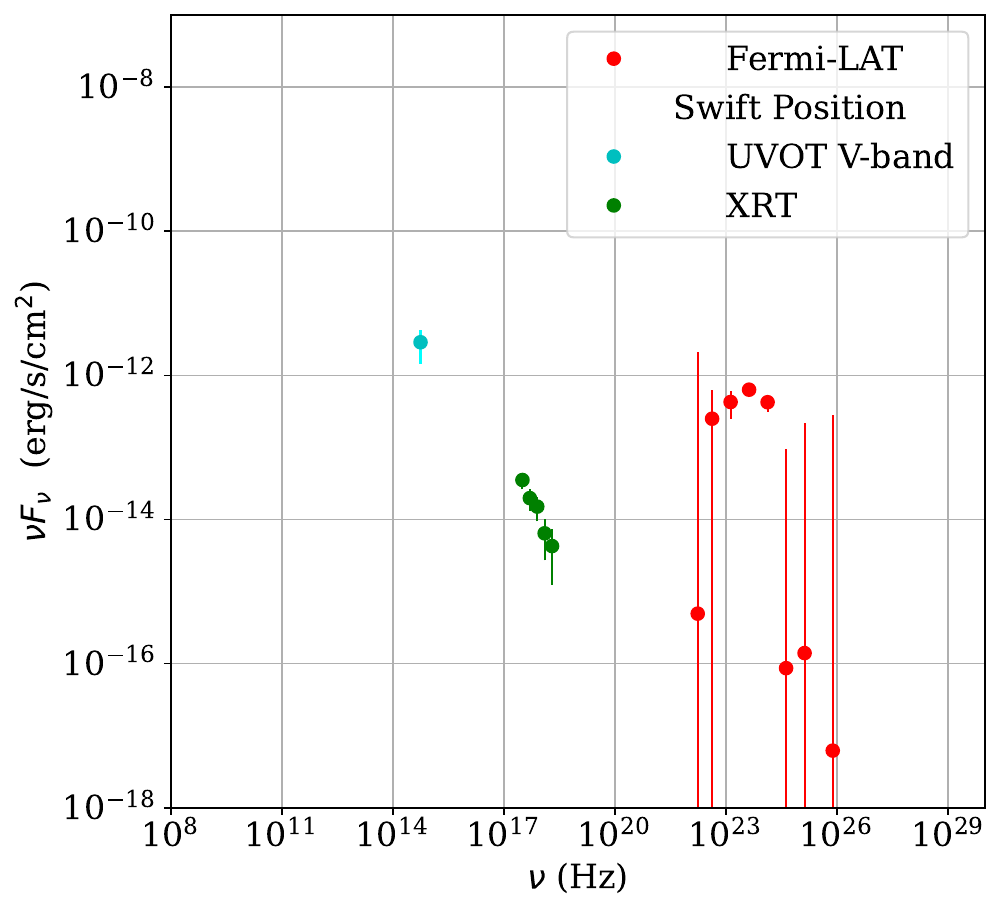}
    \caption{The gamma-ray fluxes of 4FGL J1823.8$-$3544 from the 4FGL catalog and the binned X-ray spectrum and UVOT V-band magnitude for the possible X-ray counterpart in the southern region of the uncertainty ellipse.}
    \label{fig:1823SED}
\end{figure}

Evaluating the classification of the source with the NNC while assuming association with the \textit{Fermi} gamma-ray emission, we obtain $P_{bzr} = 0.967$. This high blazar probability is probably driven by the bright $m_V$ obtained by \textit{Swift}-UVOT and suggests that this southern X-ray source, if it is associated with the gamma-ray emission observed by \textit{Fermi}-LAT, is a gamma-ray blazar. To further investigate the new X-ray source, we create a multiwavelength spectrum by plotting the \textit{Fermi} gamma-ray spectrum with the \textit{Swift}-XRT X-ray spectrum and the \textit{Swift}-UVOT magnitude for the newly discovered X-ray source, shown in Figure \ref{fig:1823SED}. 

It is possible that the \textit{Swift} source is linked to the \textit{Fermi} gamma-ray emission, in which case the UVOT, XRT, and LAT emission in Figure \ref{fig:1823SED} could evoke the characteristic two-humped spectrum of a gamma-ray blazar. Alternatively, the \textit{Swift} source may be an unrelated X-ray source not related to the \textit{Fermi} gamma-ray detection, in which case the TRAPUM pulsar is likely linked to the gamma-ray emission but is too dim in X-rays to be detected by \textit{Swift}. Coincident but unrelated X-ray sources in 4FGL ellipses in the galactic plane could be a variety of galactic and extragalactic objects, ranging from X-ray AGN to supernova remnants or X-ray active stars. Further investigation at this source is warranted including by looking for shared variability timescales or pulsations in different energy bands, and by looking for other high-energy emission in this region to better constrain the origin of the \textit{Fermi} emission.

\subsubsection{J1858.3-5424}

In the error ellipse of 4FGL J1858.3-5242 we find a $\rm{S/N} \approx 4.1$ source at the TRAPUM radio position after an extended $35\:\rm{ks}$ of \textit{Swift}-XRT observations. This X-ray detection, at the same position as the TRAPUM radio pulsar and within the 4FGL error ellipse, is a new discovery that could link the radio emission to the \textit{Fermi} gamma-ray flux. Using the $1.2 \:\rm{kpc}$ distance cited in Table 2 of \cite{TRAPUM2022} for this pulsar, we calculate an X-ray luminosity of $6.9 \times 10^{30} \:\rm{erg/s}$ for this source, which is within the typical range of X-ray luminosity for millisecond pulsars reported in \cite{Lee2018}.

Conducting power-law fitting to the X-ray data and using UVOT observations to put an upper limit on $m_V$, we find that this sources has $P_{bzr} = 0.020$, showing that the SED with the joint \textit{Fermi}, \textit{Swift}-XRT, and \textit{Swift}-UVOT data is similar to other gamma-ray pulsars. Notably, the ratio of X-ray flux to gamma-ray flux, used in \cite{Kerby2021b} as a distance-independent spectral feature, is also typical of gamma-ray pulsars. Taken together, our \textit{Swift} detection of an X-ray source provides strong evidence for linking the radio and gamma-ray emission at this location.

However, there is another X-ray source in the southern half of the 4FGL error ellipse, distinct and separate from the TRAPUM position. Given the duration of \textit{Swift}-XRT exposure and the size of the 4FGL uncertainty region, we expect approximately $0.55$ X-ray sources with $S/N > 4$ to appear purely via coincidence, so it is possible that one of the X-rays sources is an unrelated interloper after $35\:\rm{ks}$ of observations. Conducting our X-ray analysis for this alternative source, we find the $\rm{S/N} \approx 5.0$ source is brighter in X-rays than the source at the TRAPUM position and has a UVOT counterpart with $m_V = 18.75$. Using the NNC developed in \cite{Kerby2021b}, we find a blazar probability for this alternative source of $P_{bzr} = 0.007$. 

\begin{figure}
    \centering
    \includegraphics[width=\columnwidth]{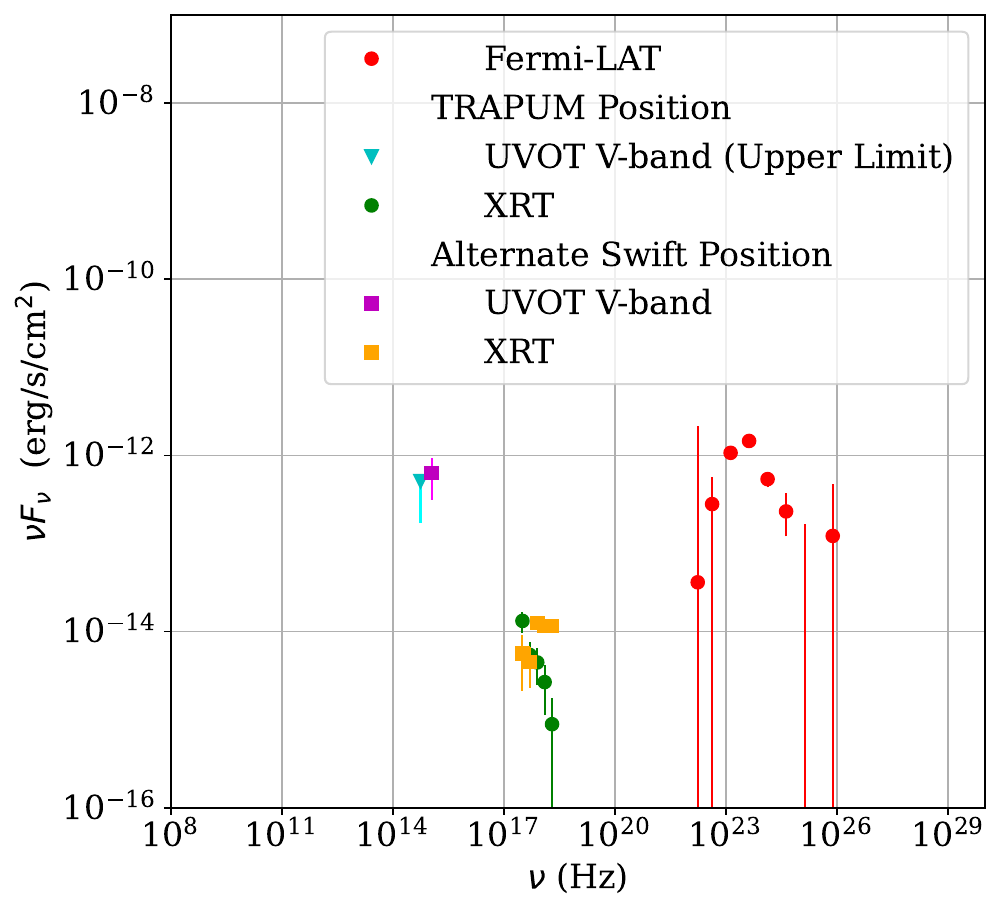}
    \caption{By comparing the \textit{Swift}-XRT spectra for the two X-ray sources in the error ellipse of 4FGL J1858.4-5242 with the \textit{Fermi}-LAT emission (red dots), we find that the TRAPUM detection (green/cyan points) and the \textit{Swift} alternative (yellow/purple points) are two possible counterparts.}
    \label{fig:1858SED}
\end{figure}

Combining the X-ray spectra of both \textit{Swift}-XRT counterparts with \textit{Swift}-UVOT magnitudes and the \textit{Fermi}-LAT emission, we create a multiwavelength spectrum of 4FGL J1858.3$-$5424 and its possible counterparts, shown in Figure \ref{fig:1858SED}. Figure \ref{fig:1858SED} shows that the UVOT/XRT emission at both the TRAPUM position and the alternate \textit{Swift} position have similar broadband properties, so it is difficult to use the SEDs alone to establish a case for either counterpart. The TRAPUM detection of a radio pulsar certainly lends credence to a link between the gamma-ray, X-ray, and radio emission at the TRAPUM position.

\subsubsection{J1906.4-1757}

At the TRAPUM position we detect no X-ray photons, with an upper limit on the count rate of $1.41 \times 10^{-3}$ counts per second (using \verb|webPIMMS| as above, $5.5 \times 10^{-14} \:\rm{erg/s/cm^2}$). Notably, the TRAPUM position lies outside of the 4FGL uncertainty ellipse. In this vein, it is worth examining the assumed link between the TRAPUM detection and the \textit{Fermi} unassociated source. Is the TRAPUM source linked to the \textit{Fermi} gamma-ray emission, or merely a coincident pulsar in the crowded galactic bulge region of the sky?

Distinct from the TRAPUM position, we detect an X-ray source in the center of the 95\% uncertainty ellipse, and we use the process described in section \ref{sec:DataAna} to conduct a power-law fit to the X-ray photons at that position. With no UVOT detection at the X-ray position, we only obtain an upper limit on the UVOT magnitude, using this value for the NNC classification. With the TRAPUM location being outside of the \textit{Fermi} ellipse and given our detection of another X-ray source within the ellipse, it is possible that our X-ray detection is the actual low-energy counterpart to the \textit{Fermi} source.

\begin{figure}
    \centering
    \includegraphics[width=\columnwidth]{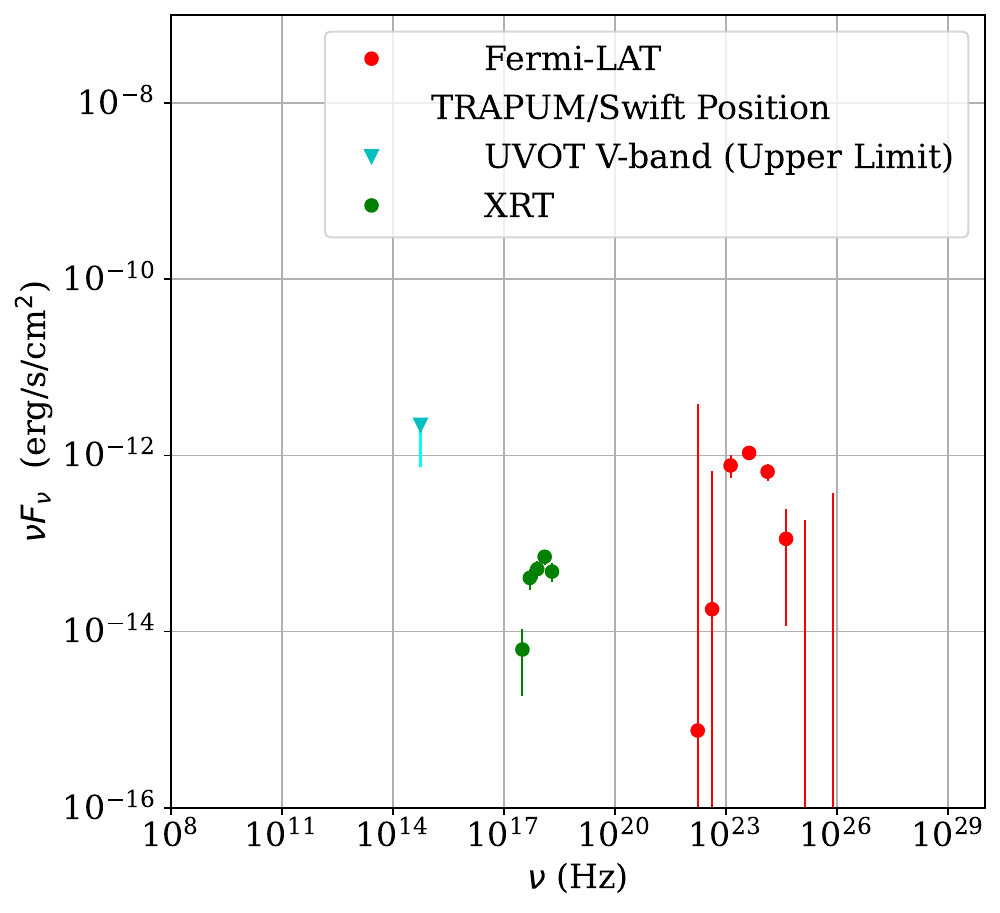}
    \caption{The gamma-ray fluxes of 4FGL J1906.4$-$1757 from the 4FGL catalogs and the binned X-ray spectrum for the detected X-ray source in the uncertainty ellipse.}
    \label{fig:1906SED}
\end{figure}

By supposing that the X-ray source is the counterpart to the \textit{Fermi} emission, we can postulate that the X-ray and gamma-ray emission are part of the same SED. Working on this assumption, the NNC described in \cite{Kerby2021b} rates this X-ray source with $P_{bzr} = 0.202$, an ambiguous rating. Similarly, the SED shown in Figure \ref{fig:1906SED} has the general shape of a gamma-ray pulsar, with keV-GeV emission in one broad hump and no UV/optical/IR components.

The likelihood of an unrelated $\rm{S/N}=7$ source within the uncertainty ellipse is very small, so it is unlikely that the detected source is an unrelated interloper. The presence of the X-ray emission with a moderately pulsar-like SED and the radial separation between the TRAPUM radio position and the \textit{Fermi} ellipse suggests that the low-energy association of 4FGL J1906.4$-$1757 is still inconclusive; the low-energy counterpart to 4FGL J1906.4$-$1757 remains an open question, with the TRAPUM radio pulsar and the pulsar-like X-ray source detected by our \textit{Swift} observations as two independent candidates.

\section{Discussion and Conclusions}
\label{sec:Coin}

\begin{figure}
    \centering
    \includegraphics[width=\columnwidth]{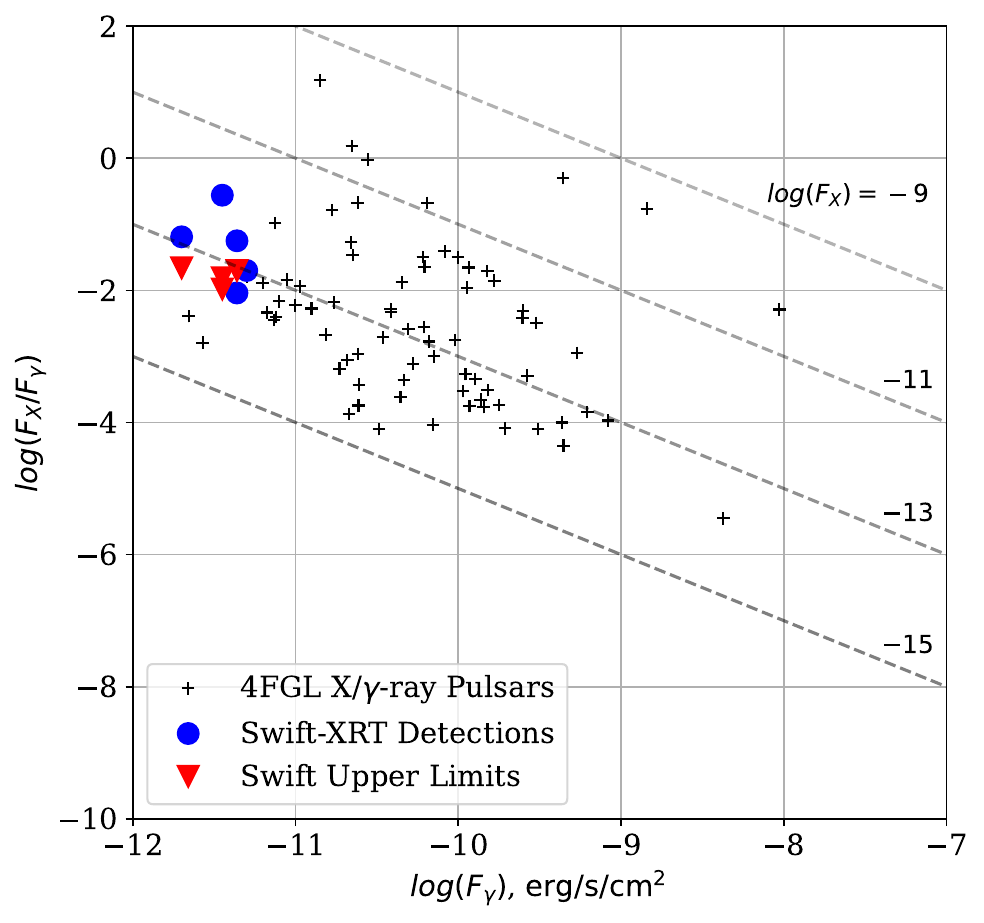}
    \caption{Comparing $F_X/F_\gamma$ and $F_\Gamma$ for the 4FGL training pulsars (black crosses), the detected \textit{Swift}-XRT counterparts (blue dots), and the TRAPUM nondetections (red triangles). Diagonal hashes show lines of constant $F_X$ in $\rm{erg/s/cm^2}$}
    \label{fig:FluxRatio}
\end{figure}

To compare our results with other X-ray/gamma-ray pulsars, we plot $F_X/F_\gamma$ versus $F_\gamma$ for the sample of 4FGL X-ray/gamma-ray pulsars used for NNC training and our sample of X-ray counterparts in Figure \ref{fig:FluxRatio}.  Creating this figure, we assign the 4FGL gamma-ray fluxes to all possible counterparts, even for those sources with spatially separate \textit{Swift} and TRAPUM detections. TRAPUM positions without \textit{Swift} detections are presented as upper limits.

Figure \ref{fig:FluxRatio} shows that the pulsars examined in this work have $F_X/F_\gamma$ flux ratios in the expected range of gamma-ray pulsars.  Similarly, TRAPUM radio pulsars without X-ray detections do not necessarily have abnormally low X-ray fluxes, in line with the predictions in Section \ref{sec:Median}.  Despite the biased sample of pulsars used for training and comparison, Figure \ref{fig:FluxRatio} shows that the pulsars discovered by TRAPUM and examined in this work are unusual only in their low gamma-ray flux, as expected for unassociated sources. Further gamma-ray surveys will probably detect additional dimmer gamma-ray pulsars and continue to expand catalogs with new sources below current flux thresholds.

Given the S/N calculations in section \ref{sec:Median}, it is not unsurprising that four of the six TRAPUM pulsars have no X-ray source detected with \textit{Swift}-XRT observations. Many brighter gamma-ray pulsars are too dim to be observed in the X-ray band. Examining \textit{Fermi} ellipses in the galactic plane with radio telescopes is a valuable step for identifying pulsars among the unassociated sources, but often \textit{Swift}-XRT observations only provide upper limits on pulsar X-ray flux. In this way, our observations at 4FGL J1623.9$-$6936 and J1757.7$-$6032 detected no X-ray sources at the TRAPUM positions or anywhere else in the \textit{Fermi} error ellipses.

At the six targets observed in this work, our \textit{Swift}-XRT and -UVOT observations have in some cases supported, extended, and confirmed the findings of the TRAPUM collaboration \citep{TRAPUM2022}, but at some targets our observations have revealed X-ray sources at separate positions in the 4FGL ellipse than the TRAPUM detection. At these targets, further investigation is warranted to solve the problem of associating the \textit{Fermi}-LAT emission with lower energy counterparts, as the detected X-ray sources could possibly be gamma-ray objects (pulsars, blazars, or other types) unrelated to the TRAPUM discovery.

Most directly, our results at 4FGL J1906.4$-$1757 warrant additional investigation and show that association of gamma-ray sources with low-energy counterparts is sometimes complex and unclear.  While the radio pulsar (PSR J1906-1754) discovery by the TRAPUM group in \cite{TRAPUM2022} is notable, there are reasons to question whether the detected pulsar is actually connected to the gamma-ray emission of 4FGL J1906.4$-$1757. First, the radio position is significantly outside of the $95\%$ uncertainty ellipse of the unassociated source from the 4FGL-DR3 catalog \citep{FermiDR32022}, as shown in Figure 2f. While it is not impossible for the point source origin of gamma-ray emission to be outside of the 95\% error ellipse, it does hint that the TRAPUM pulsar might be a coincidental discovery towards the crowded galactic bulge region. Secondly, the presence of a bright X-ray source in the \textit{Fermi} uncertainty ellipse of 4FGL J1906.4$-$1757 presents a possible alternative counterpart.

To resolve the uncertainty at J1906.4-1757 and other targets, additional observations and tests are warranted. Targeted high-energy observations in the $0.1 - 10 \:\rm{MeV}$ range could verify whether the \textit{Swift}-XRT fluxes in Figure \ref{fig:1906SED}, Figure \ref{fig:1858SED}, and Figure \ref{fig:1823SED} continue to increase and connect with the \textit{Fermi}-LAT fluxes in the gamma-ray band. This finding would be a more conclusive test of association for the possible X-ray counterpart. 

On the radio side, finding linked radio- and gamma-ray pulsations via further radio analysis at the TRAPUM pulsar could demonstrate a temporal link between the TRAPUM pulsar and the \textit{Fermi} source. Finally, upcoming eROSITA data releases should contribute to creating broadband radio-through-gamma-ray spectra of pulsars, hopefully including these six targets first observed by the TRAPUM consortium. Collecting broadband observations and stitching together expansive SEDs of pulsars can constrain emission mechanisms and expand our understanding of high-energy processes around stellar remnants.

\subsection{Summary}

At 4FGL J1623.9$-$6936 and J1757.7$-$6032 we find no X-ray source at the TRAPUM radio positions, putting upper limits on the X-ray flux from the radio pulsar. At 4FGL J1803.1$-$6708 we confirm the TRAPUM and eROSITA pulsar discovery with complementary \textit{Swift}-XRT observations.

At 4FGL J1858.3$-$5424 we find a dim X-ray source at the TRAPUM position but also discover another X-ray source that may complicate association, and at J1823.8$-$3544 and J1906.4$-$1757 we clearly detect new X-ray sources entirely separate from the TRAPUM radio positions, with new X-ray source names designated SwFX J182356.1-354800
and SwFX J190625.9-175900 respectively. The separate X-ray sources at the last three targets show that the association of the \textit{Fermi} gamma-ray source with the TRAPUM pulsars is not always straightforward. Our results at all the unassociated targets provide a detailed X-ray context to the radio discoveries of the TRAPUM consortium.

\software{Astropy \citep{Astropy}, numpy \citep{NumPy}, Matplotlib \citep{Matplotlib}, scikitlearn \citep{Scikitlearn}, FTools \citep{FTOOLS}}

\acknowledgments

This research has made use of data and/or software provided by the High Energy Astrophysics Science Archive Research Center (HEASARC), which is a service of the Astrophysics Science Division at NASA/GSFC. We gratefully acknowledge the support of NASA grants 80NSSC17K0752 and 80NSSC18K1730. Portions of this work performed at NRL were supported by NASA.

We gratefully acknowledge the work of the TRAPUM consortium in conducting radio observations and privately discussing their results with the authors.

We thank the anonymous reviewer for their helpful and insightful comments, which have greatly improved this work.

\begin{longrotatetable}

\begin{deluxetable}{cccccccccccc}
\tablecaption{Target gamma-ray/X-ray/UV/optical spectral parameters of X-ray sources at TRAPUM locations and alternate \textit{Swift} candidate counterparts. Gamma-ray properties (gamma-ray flux, spectral index, variability index, and spectral curvature) are extracted from the 4FGL-DR3 \citep{FermiDR32022} \textit{Fermi} catalog, while \textit{Swift}-XRT and -UVOT properties (S/N, X-ray flux, spectral index, V-band magnitude, and blazar probability) are found via the analysis described in \S\ref{sec:DataAna}. \label{tab:results}}

\tablewidth{\columnwidth}
\tablehead{
\colhead{Target} & \colhead{$\log(F_G)$} &\colhead{$\Gamma_G$} & \colhead{Vari.} & \colhead{Curv.}&\colhead{XRT source} & \colhead{SOSTA S/N} & \colhead{$\log(F_X)$} & \colhead{$\Gamma_X$} & \colhead{$m_V$} & \colhead{$P_{bzr}$} \\
\colhead{4FGL} & \colhead{$\rm{erg/s/cm^2}$} &\colhead{} & \colhead{} & \colhead{} & \colhead{SwXF} & \colhead{} & \colhead{$\log(\rm{erg/s/cm^2})$} & \colhead{} & \colhead{} & \colhead{}
}
\startdata
J1623.9$-$6936 & & & & & \textit{no XRT detection} & & $< -13.43$ & & & \\ \arrayrulecolor{gray} \hline 
J1757.7$-$6032 & & & & & J175745.5$-$603210 & 1.6 & $< -13.06$ & & & & \\ \hline
J1803.1$-$6708 & $-$11.30 & 2.18 & 17.45 & 5.07 & J180304.1$-$670732 & 4.25 & $-$13.00 & 1.71 & 19.85 & 0.145 \\ \hline
J1823.8$-$3544 (\textit{Swift}) & $-$11.70 & 2.31 & 15.98 & 4.19 & J182356.1$-$354800 & 4.95 & $-$12.89 & 1.95 & $\approx 17.1$ & 0.967 \\
J1823.8$-$3544 (TRAPUM) & & & & & \textit{no XRT detection} & & $< - 13.36 $ & & & \\ \hline
J1858.3$-$5424 (TRAPUM) & $-$11.36 & 2.26 & 9.44 & 5.43 & J185807.9$-$542214 & 4.17 & $-$13.40 & 2.15 & $>19.42$ & 0.028 \\
J1858.3$-$5424 (\textit{Swift}) & & & & & J185836.3$-$542709 & 5.01 & $-$12.61 & 0.16 & 18.75 & 0.007 \\ \hline
J1906.4$-$1757 (\textit{Swift}) & $-$11.45 & 2.31 & 5.52 & 3.22 & J190625.9$-$175900 & 7.58 & $-$12.01 & 0.35 & $>17.82$ & 0.202 \\
J1906.4$-$1757 (TRAPUM) & & & & & \textit{no XRT detection} & & $< -13.26$ & & & \\ \hline
\enddata
\end{deluxetable}

\end{longrotatetable}

\bibliography{main}{}

\begin{thebibliography}{}
\expandafter\ifx\csname natexlab\endcsname\relax\def\natexlab#1{#1}\fi
\providecommand{\url}[1]{\href{#1}{#1}}
\providecommand{\dodoi}[1]{doi:~\href{http://doi.org/#1}{\nolinkurl{#1}}}
\providecommand{\doeprint}[1]{\href{http://ascl.net/#1}{\nolinkurl{http://ascl.net/#1}}}
\providecommand{\doarXiv}[1]{\href{https://arxiv.org/abs/#1}{\nolinkurl{https://arxiv.org/abs/#1}}}

\bibitem[{{Abdo} {et~al.}(2013){Abdo}, {Ajello}, {Allafort}, {Baldini},
  {Ballet}, {Barbiellini}, {Baring}, {Bastieri}, {Belfiore}, {Bellazzini}, \&
  et~al.}]{Abdo2013}
{Abdo}, A.~A., {Ajello}, M., {Allafort}, A., {et~al.} 2013, \apjs, 208, 17,
  \dodoi{10.1088/0067-0049/208/2/17}

\bibitem[{Abdollahi {et~al.}(2020)Abdollahi, Acero, Ackermann, Ajello, Atwood,
  Axelsson, Baldini, Ballet, Barbiellini, Bastieri, Gonzalez, Bellazzini,
  Berretta, Bissaldi, Blandford, Bloom, Bonino, Bottacini, Brandt, Bregeon,
  Bruel, Buehler, Burnett, Buson, Cameron, Caputo, Caraveo, Casandjian, Castro,
  Cavazzuti, Charles, Chaty, Chen, Cheung, Chiaro, Ciprini, Cohen-Tanugi,
  Cominsky, Coronado-Bl{\'{a}}zquez, Costantin, Cuoco, Cutini, D'Ammando,
  DeKlotz, de~la Torre~Luque, de~Palma, Desai, Digel, Lalla, Mauro, Venere,
  Dom{\'{\i}}nguez, Dumora, Dirirsa, Fegan, Ferrara, Franckowiak, Fukazawa,
  Funk, Fusco, Gargano, Gasparrini, Giglietto, Giommi, Giordano, Giroletti,
  Glanzman, Green, Grenier, Griffin, Grondin, Grove, Guiriec, Harding, Hayashi,
  Hays, Hewitt, Horan, J{\'{o}}hannesson, Johnson, Kamae, Kerr, Kocevski,
  Kovac'evic', Kuss, Landriu, Larsson, Latronico, Lemoine-Goumard, Li,
  Liodakis, Longo, Loparco, Lott, Lovellette, Lubrano, Madejski, Maldera,
  Malyshev, Manfreda, Marchesini, Marcotulli, Mart{\'{\i}}-Devesa, Martin,
  Massaro, Mazziotta, McEnery, Mereu, Meyer, Michelson, Mirabal, Mizuno,
  Monzani, Morselli, Moskalenko, Negro, Nuss, Ojha, Omodei, Orienti, Orlando,
  Ormes, Palatiello, Paliya, Paneque, Pei, Pe{\~{n}}a-Herazo, Perkins, Persic,
  Pesce-Rollins, Petrosian, Petrov, Piron, Poon, Porter, Principe, Rain{\`{o}},
  Rando, Razzano, Razzaque, Reimer, Reimer, Remy, Reposeur, Romani, Parkinson,
  Schinzel, Serini, Sgr{\`{o}}, Siskind, Smith, Spandre, Spinelli, Strong,
  Suson, Tajima, Takahashi, Tak, Thayer, Thompson, Tibaldo, Torres, Torresi,
  Valverde, Klaveren, van Zyl, Wood, Yassine, \& Zaharijas}]{Abdollahi2020}
Abdollahi, S., Acero, F., Ackermann, M., {et~al.} 2020, The Astrophysical
  Journal Supplement Series, 247, 33, \dodoi{10.3847/1538-4365/ab6bcb}

\bibitem[{Arnaud(1996)}]{Arnaud1996}
Arnaud, K.~A. 1996, ASP Conference Series, 101

\bibitem[{{Ballet} {et~al.}(2020){Ballet}, {Burnett}, {Digel}, \&
  {Lott}}]{Ballet2020}
{Ballet}, J., {Burnett}, T.~H., {Digel}, S.~W., \& {Lott}, B. 2020, arXiv
  e-prints, arXiv:2005.11208.
\newblock \doarXiv{2005.11208}

\bibitem[{{Bessell} {et~al.}(1998){Bessell}, {Castelli}, \&
  {Plez}}]{Bessell1998}
{Bessell}, M.~S., {Castelli}, F., \& {Plez}, B. 1998, \aap, 333, 231

\bibitem[{{Blackburn}(1995)}]{FTOOLS}
{Blackburn}, J.~K. 1995, in Astronomical Society of the Pacific Conference
  Series, Vol.~77, Astronomical Data Analysis Software and Systems IV, ed.
  R.~A. {Shaw}, H.~E. {Payne}, \& J.~J.~E. {Hayes}, 367

\bibitem[{{Burrows} {et~al.}(2005){Burrows}, {Hill}, {Nousek}, {Kennea},
  {Wells}, {Osborne}, {Abbey}, {Beardmore}, {Mukerjee}, {Short}, {Chincarini},
  {Campana}, {Citterio}, {Moretti}, {Pagani}, {Tagliaferri}, {Giommi},
  {Capalbi}, {Tamburelli}, {Angelini}, {Cusumano}, {Br{\"a}uninger}, {Burkert},
  \& {Hartner}}]{Burrows2005}
{Burrows}, D.~N., {Hill}, J.~E., {Nousek}, J.~A., {et~al.} 2005, \ssr, 120,
  165, \dodoi{10.1007/s11214-005-5097-2}

\bibitem[{{Clark} {et~al.}(2022){Clark}, {Breton}, {Barr}, {Burgay},
  {Thongmeearkom}, {Nieder}, {Buchner}, {Stappers}, {Kramer}, {Becker},
  {Mayer}, {Phosrisom}, {Ashok}, {Bezuidenhout}, {Calore}, {Cognard}, {Freire},
  {Geyer}, {Grie{\ss}meier}, {Karuppusamy}, {Levin}, {Padmanabh}, {Possenti},
  {Ransom}, {Serylak}, {Venkatraman Krishnan}, {Vleeschower}, {Behrend},
  {Champion}, {Chen}, {Horn}, {Keane}, {K{\"u}nkel}, {Men}, {Ridolfi},
  {Dhillon}, {Marsh}, \& {Papa}}]{TRAPUM2022}
{Clark}, C.~J., {Breton}, R.~P., {Barr}, E.~D., {et~al.} 2022, arXiv e-prints,
  arXiv:2212.08528.
\newblock \doarXiv{2212.08528}

\bibitem[{{D'Elia} {et~al.}(2013){D'Elia}, {Perri}, {Puccetti}, {Capalbi},
  {Giommi}, {Burrows}, {Campana}, {Tagliaferri}, {Cusumano}, {Evans},
  {Gehrels}, {Kennea}, {Moretti}, {Nousek}, {Osborne}, {Romano}, \&
  {Stratta}}]{2013Delia}
{D'Elia}, V., {Perri}, M., {Puccetti}, S., {et~al.} 2013, \aap, 551, A142,
  \dodoi{10.1051/0004-6361/201220863}

\bibitem[{{Fermi-LAT collaboration} {et~al.}(2022){Fermi-LAT collaboration},
  {:}, {Abdollahi}, {Acero}, {Baldini}, {Ballet}, {Bastieri}, {Bellazzini},
  {Berenji}, {Berretta}, {Bissaldi}, {Blandford}, {Bloom}, {Bonino}, {Brill},
  {Britto}, {Bruel}, {Burnett}, {Buson}, {Cameron}, {Caputo}, {Caraveo},
  {Castro}, {Chaty}, {Cheung}, {Chiaro}, {Cibrario}, {Ciprini},
  {Coronado-Blazquez}, {Crnogorcevic}, {Cutini}, {D'Ammando}, {De Gaetano},
  {Digel}, {Di Lalla}, {Dirirsa}, {Di Venere}, {Dominguez}, {Fallah Ramazani},
  {Fegan}, {Ferrara}, {Fiori}, {Fleischhack}, {Franckowiak}, {Fukazawa},
  {Funk}, {Fusco}, {Galanti}, {Gammaldi}, {Gargano}, {Garrappa}, {Gasparrini},
  {Giacchino}, {Giglietto}, {Giordano}, {Giroletti}, {Glanzman}, {Green},
  {Grenier}, {Grondin}, {Guillemot}, {Guiriec}, {Gustafsson}, {Harding},
  {Hays}, {Hewitt}, {Horan}, {Hou}, {Johannesson}, {Karwin}, {Kayanoki},
  {Kerr}, {Kuss}, {Landriu}, {Larsson}, {Latronico}, {Lemoine-Goumard}, {Li},
  {Liodakis}, {Longo}, {Loparco}, {Lott}, {Lubrano}, {Maldera}, {Malyshev},
  {Manfreda}, {Marti-Devesa}, {Mazziotta}, {Mereu}, {Meyer}, {Michelson},
  {Mirabal}, {Mitthumsiri}, {Mizuno}, {Moiseev}, {Monzani}, {Morselli},
  {Moskalenko}, {Negro}, {Nuss}, {Omodei}, {Orienti}, {Orlando}, {Paneque},
  {Pei}, {Perkins}, {Persic}, {Pesce-Rollins}, {Petrosian}, {Pillera}, {Poon},
  {Porter}, {Principe}, {Raino}, {Rando}, {Rani}, {Razzano}, {Razzaque},
  {Reimer}, {Reimer}, {Reposeur}, {Sanchez-Conde}, {Saz Parkinson}, {Scotton},
  {Serini}, {Sgro}, {Siskind}, {Smith}, {Spandre}, {Spinelli}, {Sueoka},
  {Suson}, {Tajima}, {Tak}, {Thayer}, {Thompson}, {Torres}, {Troja},
  {Valverde}, {Wood}, \& {Zaharijas}}]{FermiDR32022}
{Fermi-LAT collaboration}, {:}, {Abdollahi}, S., {et~al.} 2022, arXiv e-prints,
  arXiv:2201.11184.
\newblock \doarXiv{2201.11184}

\bibitem[{{Gehrels} {et~al.}(2004){Gehrels}, {Chincarini}, {Giommi}, {Mason},
  {Nousek}, {Wells}, {White}, {Barthelmy}, {Burrows}, {Cominsky}, {Hurley},
  {Marshall}, {M{\'e}sz{\'a}ros}, {Roming}, {Angelini}, {Barbier}, {Belloni},
  {Campana}, {Caraveo}, {Chester}, {Citterio}, {Cline}, {Cropper}, {Cummings},
  {Dean}, {Feigelson}, {Fenimore}, {Frail}, {Fruchter}, {Garmire}, {Gendreau},
  {Ghisellini}, {Greiner}, {Hill}, {Hunsberger}, {Krimm}, {Kulkarni}, {Kumar},
  {Lebrun}, {Lloyd-Ronning}, {Markwardt}, {Mattson}, {Mushotzky}, {Norris},
  {Osborne}, {Paczynski}, {Palmer}, {Park}, {Parsons}, {Paul}, {Rees},
  {Reynolds}, {Rhoads}, {Sasseen}, {Schaefer}, {Short}, {Smale}, {Smith},
  {Stella}, {Tagliaferri}, {Takahashi}, {Tashiro}, {Townsley}, {Tueller},
  {Turner}, {Vietri}, {Voges}, {Ward}, {Willingale}, {Zerbi}, \&
  {Zhang}}]{Gehrels2004}
{Gehrels}, N., {Chincarini}, G., {Giommi}, P., {et~al.} 2004, \apj, 611, 1005,
  \dodoi{10.1086/422091}

\bibitem[{Harris {et~al.}(2020)Harris, Millman, van~der Walt, Gommers,
  Virtanen, Cournapeau, Wieser, Taylor, Berg, Smith, Kern, Picus, Hoyer, van
  Kerkwijk, Brett, Haldane, del R{\'{i}}o, Wiebe, Peterson,
  G{\'{e}}rard-Marchant, Sheppard, Reddy, Weckesser, Abbasi, Gohlke, \&
  Oliphant}]{NumPy}
Harris, C.~R., Millman, K.~J., van~der Walt, S.~J., {et~al.} 2020, Nature, 585,
  357, \dodoi{10.1038/s41586-020-2649-2}

\bibitem[{{Hunter}(2007)}]{Matplotlib}
{Hunter}, J.~D. 2007, Computing in Science and Engineering, 9, 90,
  \dodoi{10.1109/MCSE.2007.55}

\bibitem[{{Jonas}(2009)}]{Jonas2009}
{Jonas}, J.~L. 2009, IEEE Proceedings, 97, 1522,
  \dodoi{10.1109/JPROC.2009.2020713}

\bibitem[{{Kalberla} {et~al.}(2005){Kalberla}, {Burton}, {Hartmann}, {Arnal},
  {Bajaja}, {Morras}, \& {P{\"o}ppel}}]{Kalberla2005}
{Kalberla}, P.~M.~W., {Burton}, W.~B., {Hartmann}, D., {et~al.} 2005, \aap,
  440, 775, \dodoi{10.1051/0004-6361:20041864}

\bibitem[{Kaur {et~al.}(2019)Kaur, Falcone, Stroh, Kennea, \&
  Ferrara}]{Kaur2019}
Kaur, A., Falcone, A.~D., Stroh, M.~D., Kennea, J.~A., \& Ferrara, E.~C. 2019,
  \dodoi{10.3847/1538-4357/ab4ceb}

\bibitem[{{Kerby} {et~al.}(2021{\natexlab{a}}){Kerby}, {Kaur}, {Falcone},
  {Stroh}, {Ferrara}, {Kennea}, \& {Colosimo}}]{Kerby2021}
{Kerby}, S., {Kaur}, A., {Falcone}, A.~D., {et~al.} 2021{\natexlab{a}}, \aj,
  161, 154, \dodoi{10.3847/1538-3881/abda53}

\bibitem[{{Kerby} {et~al.}(2021{\natexlab{b}}){Kerby}, {Kaur}, {Falcone},
  {Eskenasy}, {Hancock}, {Stroh}, {Ferrara}, {Ray}, {Kennea}, \&
  {Grove}}]{Kerby2021b}
---. 2021{\natexlab{b}}, \apj, 923, 75, \dodoi{10.3847/1538-4357/ac2e91}

\bibitem[{Lee {et~al.}(2018)Lee, Hui, Takata, Kong, Tam, \& Cheng}]{Lee2018}
Lee, J., Hui, C.~Y., Takata, J., {et~al.} 2018, The Astrophysical Journal, 864,
  23, \dodoi{10.3847/1538-4357/aad284}

\bibitem[{Pedregosa {et~al.}(2011)Pedregosa, Varoquaux, Gramfort, Michel,
  Thirion, Grisel, Blondel, Prettenhofer, Weiss, Dubourg, Vanderplas, Passos,
  Cournapeau, Brucher, Perrot, \& Duchesnay}]{Scikitlearn}
Pedregosa, F., Varoquaux, G., Gramfort, A., {et~al.} 2011, Journal of Machine
  Learning Research, 12, 2825

\bibitem[{{Roming} {et~al.}(2005){Roming}, {Kennedy}, {Mason}, {Nousek}, {Ahr},
  {Bingham}, {Broos}, {Carter}, {Hancock}, {Huckle}, {Hunsberger}, {Kawakami},
  {Killough}, {Koch}, {McLelland}, {Smith}, {Smith}, {Soto}, {Boyd},
  {Breeveld}, {Holland}, {Ivanushkina}, {Pryzby}, {Still}, \&
  {Stock}}]{Roming2005}
{Roming}, P. W.~A., {Kennedy}, T.~E., {Mason}, K.~O., {et~al.} 2005, \ssr, 120,
  95, \dodoi{10.1007/s11214-005-5095-4}

\bibitem[{{The Astropy Collaboration} {et~al.}(2013){The Astropy
  Collaboration}, {Robitaille, Thomas P.}, {Tollerud, Erik J.}, {Greenfield,
  Perry}, {Droettboom, Michael}, {Bray, Erik}, {Aldcroft, Tom}, {Davis, Matt},
  {Ginsburg, Adam}, {Price-Whelan, Adrian M.}, {Kerzendorf, Wolfgang E.},
  {Conley, Alexander}, {Crighton, Neil}, {Barbary, Kyle}, {Muna, Demitri},
  {Ferguson, Henry}, {Grollier, Fr\'ed\'eric}, {Parikh, Madhura M.}, {Nair,
  Prasanth H.}, {G\"unther, Hans M.}, {Deil, Christoph}, {Woillez, Julien},
  {Conseil, Simon}, {Kramer, Roban}, {Turner, James E. H.}, {Singer, Leo},
  {Fox, Ryan}, {Weaver, Benjamin A.}, {Zabalza, Victor}, {Edwards, Zachary I.},
  {Azalee Bostroem, K.}, {Burke, D. J.}, {Casey, Andrew R.}, {Crawford, Steven
  M.}, {Dencheva, Nadia}, {Ely, Justin}, {Jenness, Tim}, {Labrie, Kathleen},
  {Lim, Pey Lian}, {Pierfederici, Francesco}, {Pontzen, Andrew}, {Ptak, Andy},
  {Refsdal, Brian}, {Servillat, Mathieu}, \& {Streicher, Ole}}]{Astropy}
{The Astropy Collaboration}, {Robitaille, Thomas P.}, {Tollerud, Erik J.},
  {et~al.} 2013, A\&A, 558, A33, \dodoi{10.1051/0004-6361/201322068}

\end{thebibliography}

\end{document}